\documentclass{article}
\usepackage{spconf,amsmath,graphicx,url,amssymb}
\usepackage{subfigure}
\usepackage{graphicx}
\usepackage{color}
\usepackage{enumitem}
\usepackage{units}
\usepackage{array}
\usepackage{booktabs}

\title{Modeling Beats and Downbeats with a Time-Frequency Transformer}
\name{Yun-Ning Hung$^{2,}$\sthanks{The author conducted this work as an intern at ByteDance.}, Ju-Chiang Wang$^{1}$, Xuchen Song$^{1}$, Wei-Tsung Lu$^{1}$, and Minz Won$^{1}$}
\address{$^{1}$ ByteDance, Mountain View, CA, USA\\ 
$^{2}$ Center for Music Technology, Georgia Institute of Technology, Atlanta, GA, USA \\
{\small\tt amy.hung@gatech.edu, \{ju-chiang.wang, xuchen.song, weitsung.lu, minzwon\}@bytedance.com}}

\begin{document}
\ninept
\maketitle
\begin{abstract}
Transformer is a successful deep neural network (DNN) architecture that has shown its versatility not only in natural language processing but also in music information retrieval (MIR). In this paper, we present a novel Transformer-based approach to tackle beat and downbeat tracking. This approach employs SpecTNT (Spectral-Temporal Transformer in Transformer), a variant of Transformer that models both spectral and temporal dimensions of a time-frequency input of music audio. A SpecTNT model uses a stack of blocks, where each consists of two levels of Transformer encoders. The lower-level (or spectral) encoder handles the spectral features and enables the model to pay attention to harmonic components of each frame. Since downbeats indicate bar boundaries and are often accompanied by harmonic changes, this step may help downbeat modeling. The upper-level (or temporal) encoder aggregates useful local spectral information to pay attention to beat/downbeat positions. We also propose an architecture that combines SpecTNT with a state-of-the-art model, Temporal Convolutional Networks (TCN), to further improve the performance. Extensive experiments demonstrate that our approach can significantly outperform TCN in downbeat tracking while maintaining comparable result in beat tracking.
\end{abstract}



\section{Introduction}\label{sec:introduction}

Beat is regarded as one of the fundamental rhythmic units perceived by humans, and a downbeat is conceived to occur at the first beat of a bar (measure), which usually indicates the start of a chord. 
Beat and downbeat tracking is concerned with developing algorithms to detect the beats and downbeats as pulse signals in music audio. This is a well-defined problem in music information retrieval (MIR) and has been a regular task in MIREX challenge \footnote{https://www.music-ir.org/mirex/wiki/MIREX\_HOME} for over a decade. 

In recent years, researchers proposed to employ deep neural network (DNN)-based methods to model beats and downbeats, which are regularly repeating events in the audio sequence. B{\"o}ck et al. \cite{bock2011enhanced} successfully used a recurrent neural network (RNN) with input mel spectrogram features to jointly model beats and downbeats. Such a system requires annotated data to train and then predict accordingly. This has set up a data-driven paradigm for the subsequent systems, where improvement can be expected by either involving more training data or developing a more advanced machine learning algorithm. 

Compared to beats, predicting downbeats still remains a challenge, because it requires a more complicated awareness of harmonic, timbral, and structural context, which may be related to chords, instrumentation, and percussive arrangements \cite{durand2016robust}. To better model these aspects, one need to explore more local spectral information, and at the same time allow the important local information to be exchangeable among different temporal positions far apart in the audio sequence, since a measure may require a longer audio context to be discovered.  

Recently, Transformer architecture \cite{vaswani2017attention, devlin2018bert} has brought lots of attention due to its remarkable performance in natural language processing. For song-level classification tasks in MIR, researchers treat it as a temporal encoder to aggregate the temporal features to represent a music audio sequence \cite{won2019toward, won2021transformer}. As a replacement of RNN, it also works well for chord recognition \cite{chen2019harmony,park2019bi}.
However, these prior works did not attempt to make full exploitation of Transformer to model the spectral information that interacts with the temporal encoder as a whole.
On the other hand, Transformer is data-hungry in nature. For many MIR tasks, annotations are very expensive, and the size of annotated data is typically small. Therefore, training an effective Transformer-based model without modification and data augmentation is likely infeasible. 

In this paper, we propose to use SpecTNT (Spectral-Temporal Transformer in Transformer) \cite{specTNT}, a variant of Transformer that models the spectrogram along both time and frequency axes. To the best of our knowledge, this is the first successful attempt to apply Transformer-based model to beat and downbeat tracking.
The basic principle of SpecTNT stems from the interaction between two levels of Transformer encoders, namely \emph{spectral Transformer} and \emph{temporal Transformer}. The former is responsible for extracting the spectral features for each frame. Whereas the latter is capable of exchanging the local information over the time axis. Thanks to the hierarchical design \cite{specTNT}, SpecTNT permits a smaller number of parameters compared to the original Transformer \cite{devlin2018bert}, so it can maintain good generalization ability for MIR tasks with a smaller size of training data. To further improve the performance, we also propose a novel structure that combines SpecTNT with a state-of-the-art (SOTA) model, temporal convolutional networks (TCN). We evaluate our proposed systems on various public datasets. Our results demonstrate that SpecTNT can achieve SOTA performance in downbeat tracking on most of the datasets while maintaining comparable performance in beat tracking. The combined structure of SpecTNT with TCN can further boost the performance.



\section{Related Work}

Early approaches in beat tracking mostly focused on incorporating heuristic information into the systems. For example, Dixon et al. \cite{dixon2001automatic} proposed to estimate the tempo and beats by detecting salient rhythmic events at various metrical levels. Klapuri et al. \cite{klapuri2005analysis} proposed a probabilistic model to process the music accents and estimate beat patterns. On the other hand, musical events such as chord changes and drum patterns were explored to facilitate the analysis of beat structures \cite{goto2001audio}. A two-stage approach was also proposed by \cite{davies2007context} to estimate tempo and beat sequentially, where the process attempted to replicate human ability to tap along music. 

More recently, data-driven approaches 
have been widely used \cite{bock2011enhanced}, since they enable a system to automatically learn the relevant information to detect beats from annotated audio. Specifically, DNN architectures such as RNN \cite{bock2016madmom} and TCN \cite{matthewdavies2019temporal} were employed to effectively model the long sequence of beat events from audio. Multi-task approaches were also introduced to include more types of annotations in the learning system, such as tempo and downbeat \cite{bock2019multi, bock2020deconstruct, bock2016joint, fuentes2019music}, owing to their high relevance to beats, and each task was improved by training multiple tasks 
altogether. These endeavors have improved beat tracking significantly, and the evaluation scores can easily achieve over 90\% on many benchmarks.

As noted in the previous section, downbeat tracking still remains a challenge. Several prior instances tackled this problem by using musical features such as time-signature \cite{krebs2013rhythmic}, onset positions \cite{jehan2005downbeat}, chords \cite{papadopoulos2010joint}, and spectral difference in beats \cite{davies2006spectral}. With machine learning approaches, we believe it can be improved by introducing a novel model architecture with more training data.

\section{Proposed Method}\label{sec:method}

\subsection{System Overview} \label{sec:overview}

\begin{figure}[t]
  \centering
  \includegraphics[width=\columnwidth]{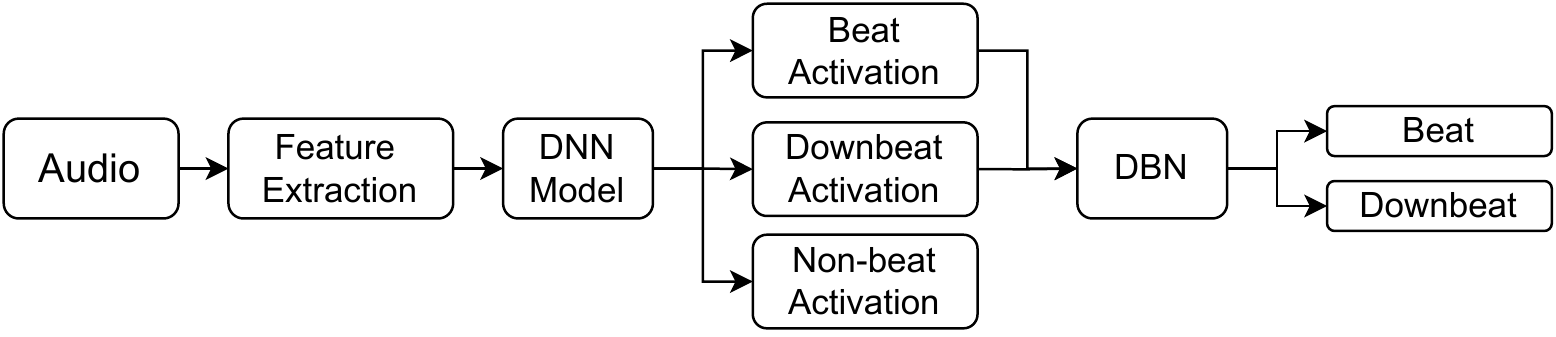}
  \caption{Pipeline for the beat and downbeat tracking system.}
  \label{fig:pipeline}
\end{figure}

The system pipeline follows the standard setting \cite{bock2016madmom, bock2020deconstruct} as depicted in Figure~\ref{fig:pipeline}.
In this work, our contribution is to introduce two novel DNN models in the pipeline, and the rest of the modules remain as conventional. For feature extraction, we apply trainable harmonic filters to convert the input audio into a harmonic representation \cite{won2020data}. This step helps capture useful harmonic information and has proven beneficial in music auto-tagging \cite{won2020data}. Then, we define the DNN model to contain three modules: front-end, main model, and a linear layer. The DNN model predicts the likelihoods of beat, downbeat, and non-beat at each time step and forms three corresponding activation functions. Finally, a dynamic Bayesian network (DBN) \cite{bock2014multi} decodes the activation functions to output the timestamps of beats and downbeats. In what follows, we will detail three options of DNN models (two proposed and one baseline models) used in this work.


\begin{figure}
\centering  
    \begin{minipage}[b]{0.23\linewidth}
        \centering
        \subfigure[SpecTNT]{\label{fig:specTNT}\includegraphics[width=\textwidth]{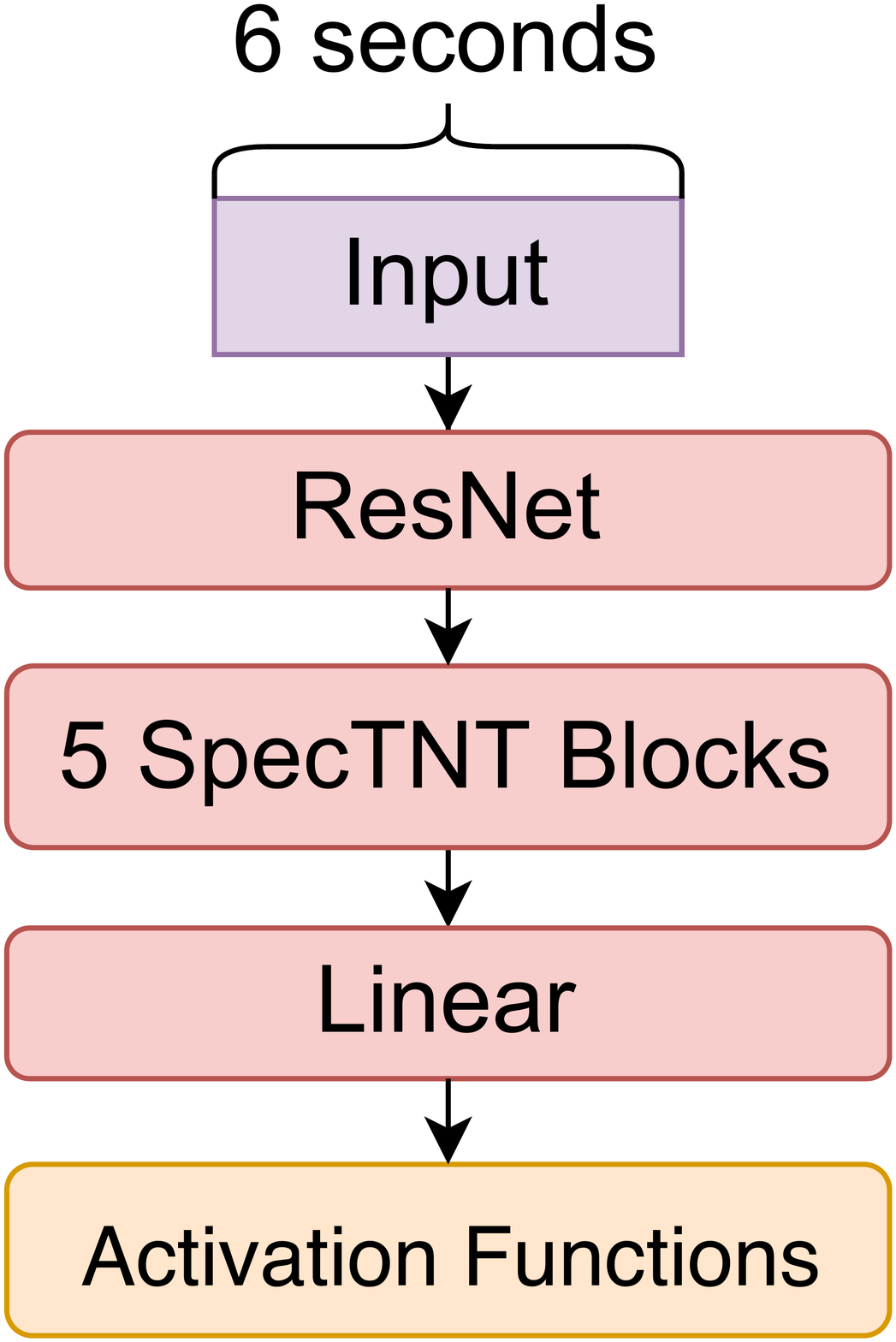}\par\vspace{5pt}}
        \subfigure[TCN]{\label{fig:tcn}\includegraphics[width=\textwidth]{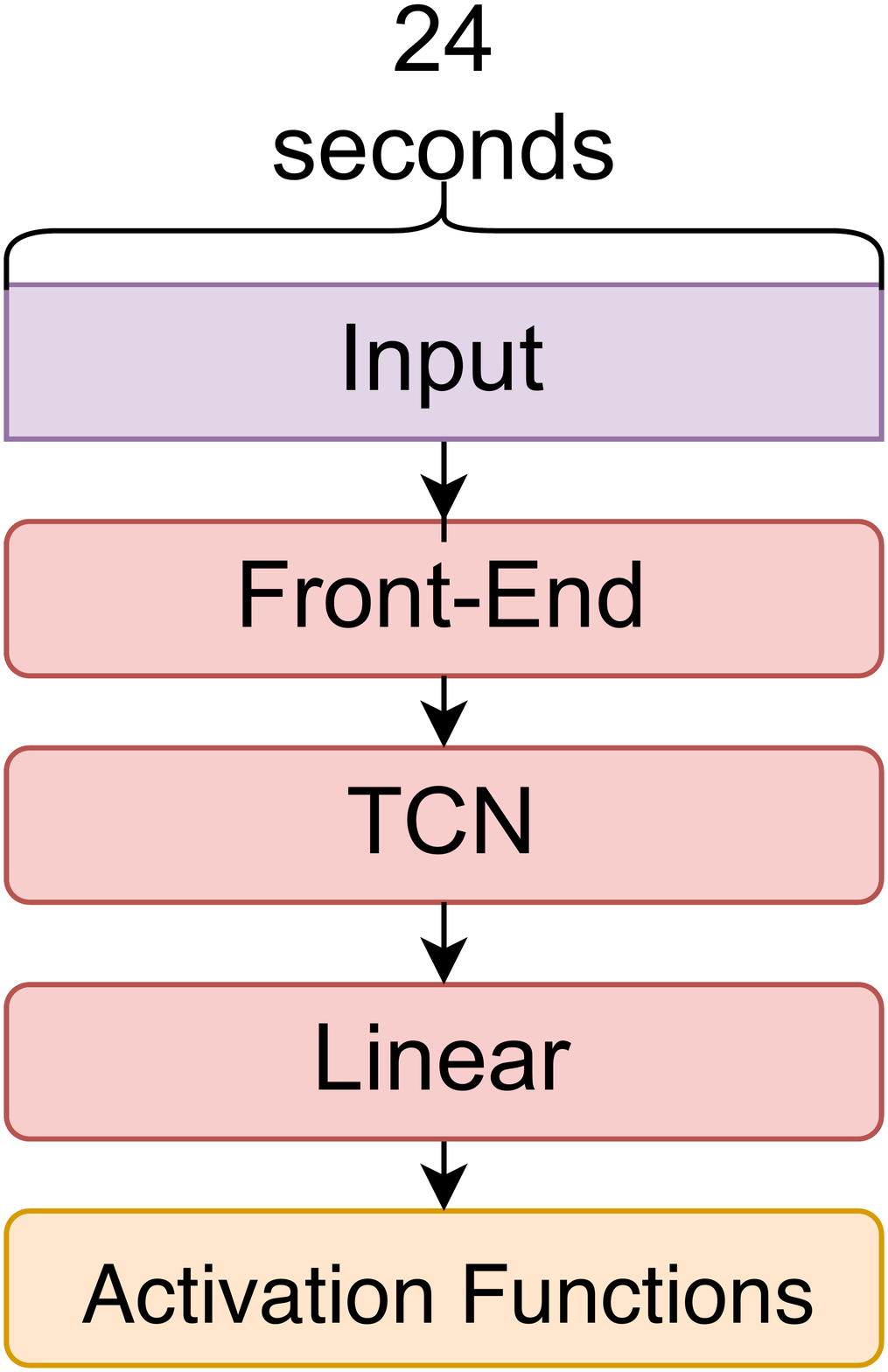}}
    \end{minipage}
    \hspace{4.00mm}
    \begin{minipage}[b]{0.57\linewidth}
        \centering
        \subfigure[SpecTNT-TCN]{\label{fig:specTNT_tcn}\includegraphics[width=0.9\textwidth]{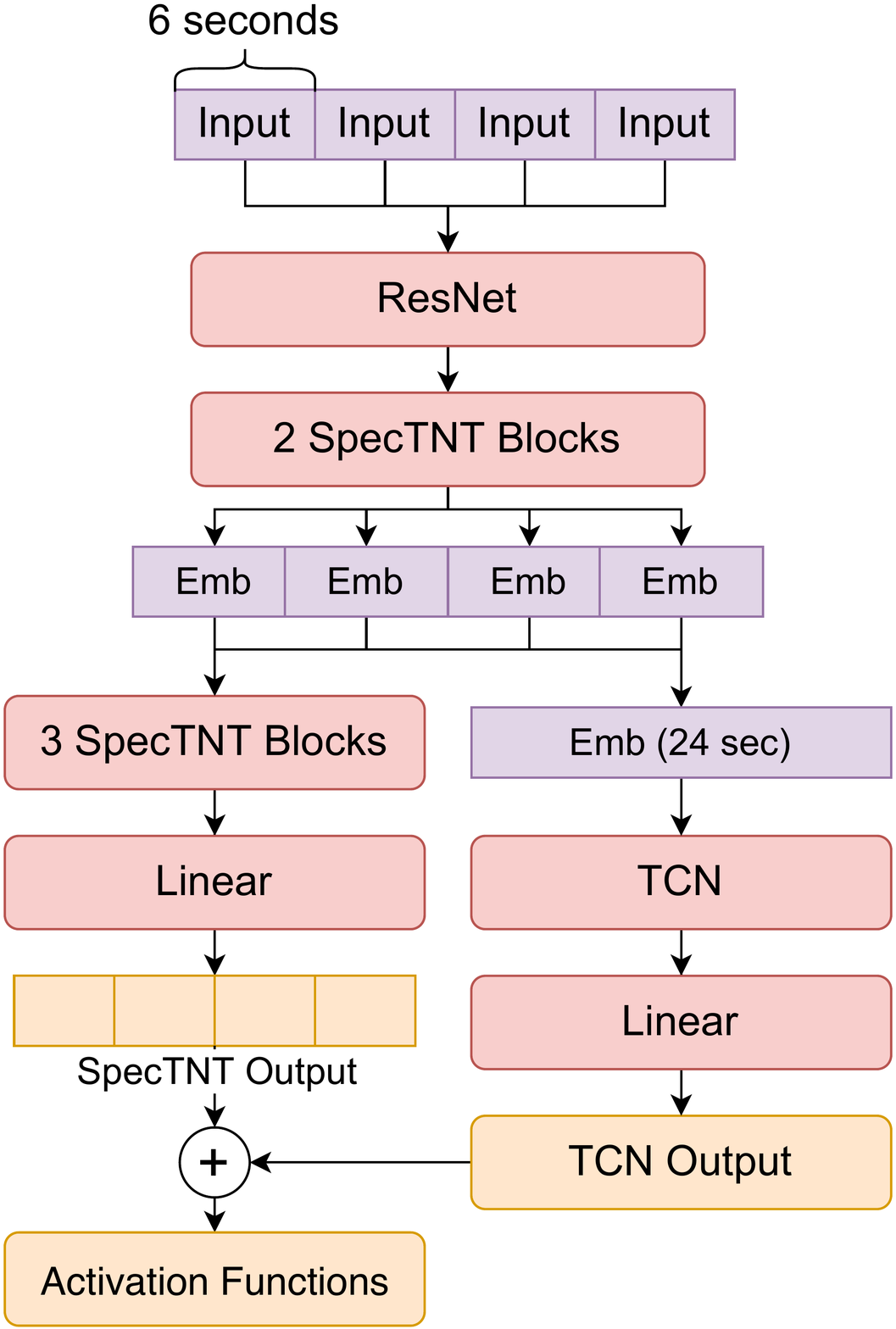}}
    \end{minipage}

\caption{Model architecture overview.}
\label{fig:dnn_models}
\end{figure}

\subsection{Spectral-Temporal Transformer in Transformer (SpecTNT)} \label{sec:spectnt}

An input harmonic representation is first processed by the front-end convolutional layers (called \emph{ResNet} \cite{he2016identity}). Then, we stack 3 residual units, each uses 256 feature maps and a kernel size of 3. This design is similar to that proposed in \cite{won2021transformer}.

The main SpecTNT module (see Fig. \ref{fig:specTNT}) 
is formed by stacking multiple SpecTNT blocks. A SpecTNT block consists of a \emph{spectral encoder} and a \emph{temporal encoder}. We use 64 feature maps with 4 attention heads for the spectral encoder to extract the spectral features into a set of \emph{frequency class tokens} (FCTs) at each time step. Through the attention mechanism, a FCT can characterize useful harmonic and timbral components, which may better represent a chord or instrumentation. For the temporal encoder, we use 256 feature maps and 8 attention heads. It enables the important local spectral information to be exchangeable via FCTs to pay attention to the beat/downbeat positions. 
More details can be found in \cite{specTNT}. Finally, we stack 5 SpecTNT blocks, followed by a linear layer to output three activation functions for beat, downbeat, and non-beat.



\subsection{Baseline TCN} 
\label{sec:tcn}

We adopt the TCN architecture \cite{bock2020deconstruct, matthewdavies2019temporal} with slight modifications as our baseline model (see Fig. \ref{fig:tcn}), since it is regarded as the current SOTA. We re-implemented the model following \cite{bock2020deconstruct}. 
Readers can refer to the codes recently released by the authors in \cite{tempobeatdownbeat:book}.

The TCN front-end is composed of three convolutional layers, each having a filter size of 20 and kernels with various sizes as described in \cite{bock2020deconstruct}. Each convolutional layer also applies a max-pooling to downsample the frequency dimension.
After the front-end, the main model contains 11 dilated convolutional layers. Each convolutional layer has two dilated convolution blocks, and each block has a twice the dilation rate of the another. This design has resulted in better performance according to \cite{bock2020deconstruct}. Finally, a linear layer is used to output the three activation functions.

\subsection{Combining SpecTNT and TCN} \label{sec: spectnt_tcn}

SpecTNT and TCN are fundamentally different in many aspects, so we believe they can be complementary to each other if they are combined. From our pilot study, we also found that modeling beats and downbeats jointly can result in a sub-optimal model for either beat or downbeat tracking. Since the amount of downbeat annotations is less than that of beat annotations, downbeat optimum will require more epochs to attain. However, longer training may lead to worse performance for beat, because the model over-fits the beat annotations. We will discuss this later in Section \ref{sec:discussion}.
To this end, we propose a fused architecture, called \emph{SpecTNT-TCN} (see Fig. \ref{fig:specTNT_tcn}). We hope it can gain more flexibility during training so as to facilitate a joint optimum for beat and downbeat tracking simultaneously.

Specifically, SpecTNT-TCN first uses 2 SpecTNT blocks at the front, and then arranges a two-stream structure that allocates a TCN branch and 3 succeeding SpecTNT blocks in parallel.
Because SpecTNT requires shorter input, as discussed in Section \ref{sec:implementation}, we divide the input into smaller chunks at the SpecTNT front-end.
The two branches have their own independent linear layers, and the final output is the average of the two branch outputs. The TCN branch follows the same TCN module described in Section \ref{sec:tcn}. To sum up, during training, parameters of each branch are updated according to its own loss and the same SpecTNT parameters at the front are updated according to the added losses. Final outputs are the average of the outputs from each branch.
From our observations, such design is better than directly merging the outputs of two full models (i.e., averaging the two outputs of Fig. \ref{fig:specTNT} and Fig. \ref{fig:tcn}).


\section{Experiments}\label{sec:experiment}

\subsection{Implementation Details}
\label{sec:implementation}

For data augmentation, we adopt a similar method to that proposed in \cite{bock2020deconstruct} for training. Specifically, the hop size to construct the STFT representation is randomly changed during training using a normal distribution with the estimated tempo from the training data as the mean and $5\%$ standard deviation. This can result in more samples of various tempi. 
Then, multiple triangular filters are applied to STFT to obtain the harmonic representation.
Following \cite{bock2020deconstruct}, we apply the same target widening strategy for the frame-wise beat and downbeat labels, where the neighbouring frames of an annotated frame are set to be positive, but with a lower weight of 0.5. We use a weighted binary cross-entropy to compare between each target and prediction.

Based on our pilot study, we determine the input length to be 6 seconds for SpecTNT and 24 seconds for TCN. 
Although several counterpart systems (e.g., \cite{bock2016joint} and \cite{bock2020deconstruct}) use 6-second, we found 24-second works better for TCN.
In SpecTNT-TCN, the input is 24-second, where it is later divided into four 6-second chunks for the SpecTNT front module (see Fig. \ref{fig:specTNT_tcn}).
For STFT, we use sampling rate of 16000 Hz and a window length of 1024.

We apply random sampling to load a random chunk into a batch, instead of sequentially loading a (6- or 24-second) chunk from the beginning of each training song. That is, we first enumerate all the valid training chunks into a list by using a sliding window (6- or 24-second) with 1-second hop size on every song, and the last chunk of a song shall not exceed the length of the song. Then, the data loader draws a sample uniformly from the list during training. Therefore, a batch can include chunks from different locations of different songs. This technique leads to faster and better convergence.



We use PyTorch 1.8 and the Adam optimizer \cite{kingma2014adam} with 0.001 learning rate and 80\% weight decay.
The numbers of parameters for TCN, SpecTNT, and SpecTNT-TCN are 86,679, 4,637,392, and 4,692,896, respectively.
We train and test the models using 4 Tesla-V100-SXM2-32GB GPUs with batch sizes of 128, 128, and 32 for SpecTNT, TCN, and SpecTNT-TCN, respectively. Each epoch runs 500 mini-batches of training before a validation. 
Using the aforementioned specifications, TCN, SpecTNT, and SpecTNT-TCN take about 5.8 minutes, 10 minutes, and 6.5 minutes, respectively, to train 500 batches, and 6 seconds, 12.6 seconds, and 15.1 seconds, respectively, to inference 100 30-second audio clips. 
The best model is selected based on the validation set for testing. 

\subsection{Experiment Setting}

We use nine public datasets to evaluate the proposed methods. Since the beat/downbeat tracking datasets are commonly small in size, we tried to collect as many as possible and combined them to conduct training and evaluation in an 8-fold cross-validation manner following \cite{bock2020deconstruct}. These datasets include \emph{Beatles} \cite{davies2009evaluation}, \emph{Ballroom} \cite{gouyon2005computational}, \emph{SMC} \cite{gouyon2005computational}, \emph{Hainsworth} \cite{hainsworth2004particle}, \emph{Simac} \cite{holzapfel2012selective}, \emph{HJDB} \cite{hockman2012one}, \emph{RWC-Popular} \cite{goto2002rwc}, and \emph{Harmonix Set} \cite{nieto2019harmonix}. An additional dataset, \emph{GTZAN} \cite{marchand2015swing}, is used for testing only. 

We note that the \emph{Harmonix Set} is comparably larger than other datasets, containing 912 pop/rock songs with more than 56 hours of beat- and downbeat-annotated audio content. Our copy is slightly different from the original one in \cite{nieto2019harmonix}. 
The original audio files corresponding to the annotations are not available, but reference YouTube links are provided. 
Our investigation indicated there are many errors when aligning the annotations with the YouTube audio, so we searched for the audio 
and manually adjusted the annotations to ensure the labels and timestamps are sensible and aligned to the audio.


For evaluation, we report the standard metrics: \emph{F1}, \emph{CMLt}, and \emph{AMLt}, with a tolerance window of 70 ms as defined in \cite{davies2009evaluation}. The resulting scores are computed using the \texttt{\footnotesize mir\_eval} package \cite{raffel2014mir_eval}. 

\begin{figure}
\centering  
\subfigure[Spectral attention]{\label{fig:spectral}\includegraphics[width=0.44\columnwidth]{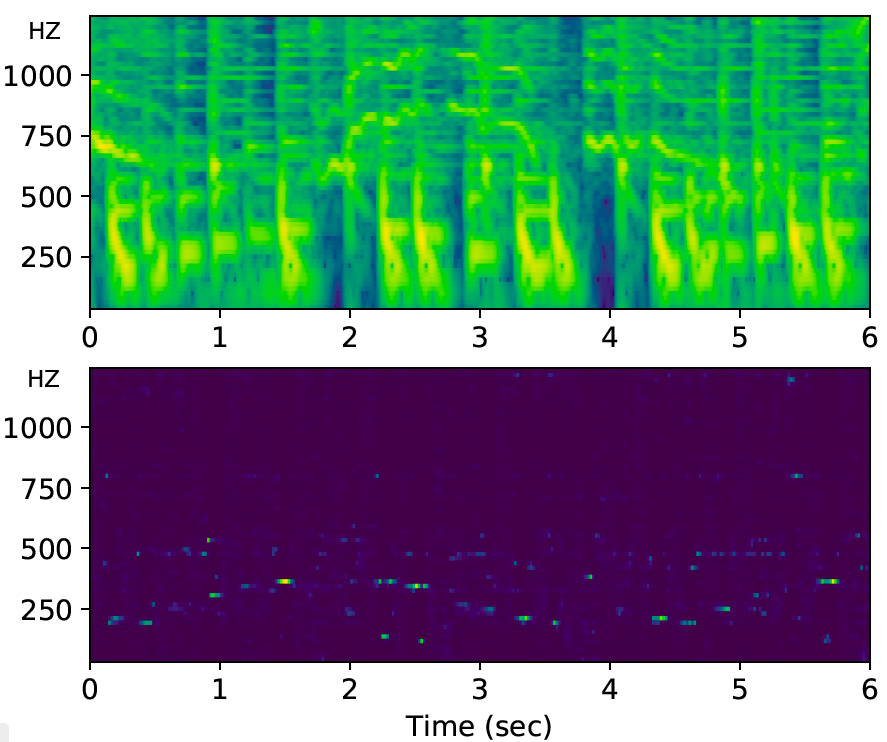}}
\subfigure[Temporal attention]{\label{fig:temporal}\includegraphics[width=0.55\columnwidth]{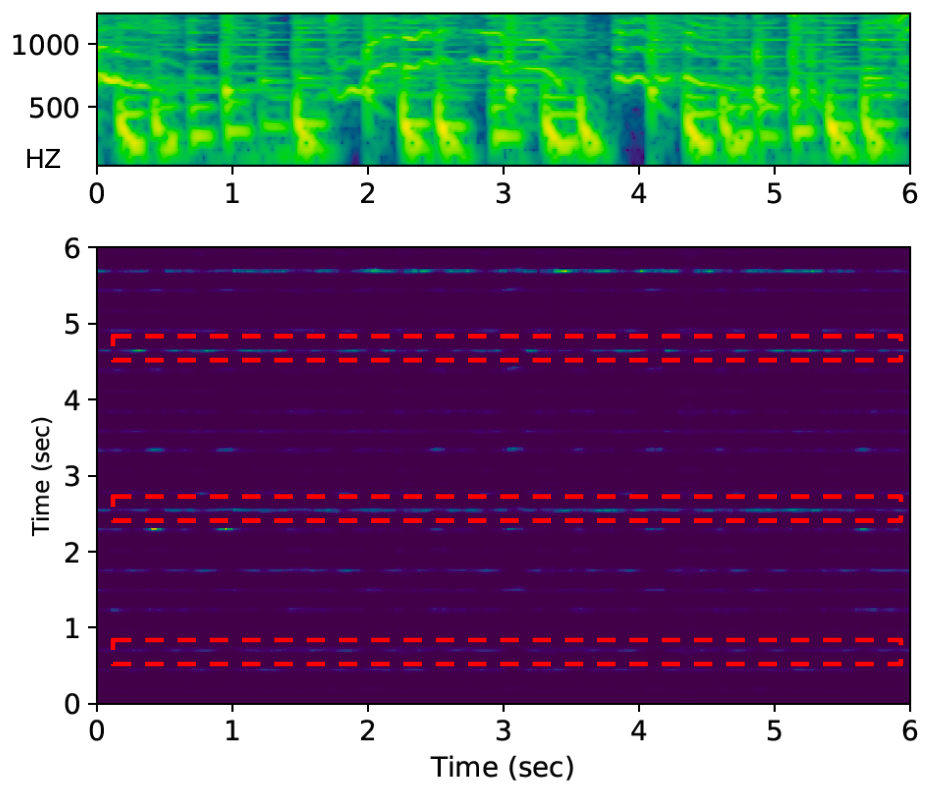}}
\caption{Visualization of the attention maps for a \emph{GTZAN} example, ``pop.00098.wav'' from 6 to 12 seconds. Upper subplot displays one of the time-synchronized harmonic representations.}
\end{figure}

\subsection{Attention Map Visualization}

To better understand the attention mechanism in SpecTNT when modeling beats and downbeats, we visualize one of the attention heads of the last SpecTNT block during the inference. For the spectral attention (see Fig. \ref{fig:spectral}), the frequency attention scores that generate the FCT of each frame are obtained. It can be seen that the attention map successfully characterizes the harmonic components. The spectral attention map emphasizes on the bass notes between 200 and 400 Hz which matches the bass line contour in the harmonic representation. For the temporal attention (see Fig. \ref{fig:temporal}), we display the time-by-time self-attention map from one of the 8 attention heads that we consider mostly relevant to the downbeat modeling. The red dashed rectangles highlight the downbeat locations. We can see that almost every frame attends to the downbeat locations, i.e., each column vector (a frame) has higher scores at the downbeat times. This will help generate the output embeddings that emphasize the downbeats, and predict higher probabilities for them accordingly.

\subsection{Results and Discussion}
\label{sec:discussion}

The beat and downbeat tracking results are presented in Table~\ref{table:beat} and Table~\ref{table:downbeat}, respectively. 
We include the results from prior literature for reference, although they are not necessarily comparable to ours, as the training data and experimental settings are different.
The performance scores of \emph{Harmonix Set} for B{\"o}ck et al.~\cite{bock2016joint} were evaluated on the predictions made by the madmom package \cite{bock2016madmom}, which uses the method proposed in \cite{bock2016joint}. \emph{Harmonix Set} was not included in the training set of madmom.  
 

Comparing the results among SpecTNT, TCN, and SpecTNT-TCN in Table~\ref{table:beat}, we observe that SpecTNT is generally better than TCN, and that SpecTNT-TCN outperforms SpecTNT in most cases. This demonstrates that SpecTNT is a promising DNN architecture for this task. 
In particular, SpecTNT and SpecTNT-TCN perform significantly better on \emph{SMC}, one of the most challenging datasets in beat tracking. Our qualitative inspections indicate that SpecTNT is superior in handling expressive timing.
For instance, it performs better in cases such as ``SMC\_00111'' (with a pause) and ``SMC\_00165'' (with varying tempo).
However, it may suffer from non-trivial percussive patterns that yield phase errors (e.g., ``SMC\_00213''). Other than that, the results also show that SpecTNT seems less successful on \emph{Ballroom} and \emph{Hainsworth}. This could be attributed to the sub-optimal issue mentioned in Section \ref{sec: spectnt_tcn}. Further investigations reveal that SpecTNT tends to over-fit beat annotations of these two datasets early during training, and then the performance drops slightly in the later epochs.
This is expected, because \emph{Hainsworth} covers more diverse genres (e.g., classical, folk, and jazz), and \emph{Ballroom} is compiled with old-school dance music. These two datasets provide relatively less examples for their specific genres, as compared to a much larger dataset, \emph{Harnmonix set}, which offers fairly more examples of pop-style in the training set.
Since our model selection criterion considers the joint performance of beat and downbeat, a model that performs much better in downbeat than in beat is picked eventually (and it requires more epochs). Nevertheless, SpecTNT-TCN is more robust to this issue and shows improvements over SpecTNT and TCN. 



\begin{table}[t]
\centering
\resizebox{\columnwidth}{!}{
\begin{tabular}{l|ccc|ccc} 
\toprule
 & F1 & CMLt & AMLt & F1 & CMLt & AMLt  \\
 \midrule
 & \multicolumn{3}{c|}{\textit{RWC-POP}} & \multicolumn{3}{c}{\textit{Harmonix Set}} \\
B{\"o}ck et al. \cite{bock2016joint} & .943 & - & - & .933$^\dagger$ & .841$^\dagger$ & .938$^\dagger$ \\
TCN (baseline)& .947 & .922 & .952 & .946~~ & .895~~ & .942~~  \\
SpecTNT       & .953 & .925 & .957 & .947~~ & .896~~ & .943~~  \\
SpecTNT-TCN   & .950 & .925 & .958 & .953~~ & .939~~ & .959~~~  \\ [0.2cm]

 & \multicolumn{3}{c|}{\textit{SMC }} & \multicolumn{3}{c}{\textit{Beatles}} \\
B{\"o}ck et al. \cite{bock2016joint} & .516~~ & .406~~ & .575 & .918 & - & - \\
B{\"o}ck et al. \cite{bock2020deconstruct}  & .544~~ & .443~~ & .635 & - & - & - \\ 
TCN (baseline) & .560~~ & .474~~ & .621 & .933 & .870 & .933  \\
SpecTNT        & .602$^\star$ & .515$^\star$ & .661 & .940 & .898 & .929  \\
SpecTNT-TCN    & .605$^\star$ & .514$^\star$ & .663 & .943 & .896 & .938  \\  [0.2cm]

 & \multicolumn{3}{c|}{\textit{Ballroom}} & \multicolumn{3}{c}{\textit{Hainsworth}} \\
Davies et al. \cite{matthewdavies2019temporal} & .933~~  & .881~~  & .929  & .874  & .795  & .930 \\
B{\"o}ck et al. \cite{bock2020deconstruct}  & .962~~  & .947~~  & .961   & .902  & .848  & .930 \\ 
TCN (baseline) & .940~~  & .870~~  & .957  & .860  & .849  & .915  \\
SpecTNT        & .927~~  & .856~~  & .939  & .866  & .865  & .914  \\
SpecTNT-TCN  & .962$^\star$  & .939$^\star$  & .967 & .877  & .862  & .915   \\

\bottomrule
 \end{tabular}}
\caption{Beat tracking results using an 8-fold cross validation, where $^\star$ denotes statistical significance compared to TCN (baseline), and $^\dagger$ denotes the predictions made by madmom \cite{bock2016madmom}.}

\label{table:beat}
\end{table}

\begin{table}[t]
\normalsize
\centering
\resizebox{\columnwidth}{!}{
\begin{tabular}{l|ccc|ccc} 
\toprule
 & F1 & CMLt & AMLt & F1 & CMLt & AMLt  \\
 \midrule
 & \multicolumn{3}{c|}{\textit{RWC-POP}} & \multicolumn{3}{c}{\textit{Harmonix Set}} \\
B{\"o}ck et al. \cite{bock2016joint} & .861 & - & - & .804$^\dagger$ & .747$^\dagger$ & .873$^\dagger$ \\ 
TCN (baseline)& .930 & .928 & .938 & .873~~ & .839~~ & .908~~  \\
SpecTNT       & .941 & .929 & .957 & .897$^\star$  & .862~~  & .924$^\star$  \\
SpecTNT-TCN   & .945 & .939 & .959 & .908$^\star$  & .872$^\star$  & .928$^\star$ \\
[0.2cm]

 & \multicolumn{3}{c|}{\textit{Ballroom}} & \multicolumn{3}{c}{\textit{Beatles }} \\
Fuentes et al. \cite{fuentes2018analysis}  & .830~~ & - & - & .860 & - & - \\
B{\"o}ck et al. \cite{bock2020deconstruct}  & .916~~ & .913~~ & .960 & .837 & .742~~ & .862 \\
TCN (baseline) & .841~~ & .788~~ & .937 & .843 & .767~~ & .851  \\
SpecTNT        & .884$^\star$  & .835$^\star$  & .937 & .867 & .810$^\star$  & .860  \\
SpecTNT-TCN    & .937$^\star$  & .927$^\star$  & .968 & .870 & .812$^\star$ & .865 \\
[0.2cm]

 & \multicolumn{3}{c|}{\textit{Hainsworth}} & \multicolumn{3}{c}{\textit{}} \\
B{\"o}ck et al. \cite{bock2020deconstruct}  & .722~~ & .696~~ & .872  &  &  &    \\
TCN (baseline) & .682~~ & .683~~ & .852 &  & &   \\
SpecTNT        & .729$^\star$  & .734$^\star$  & .879 &  & &   \\
SpecTNT-TCN    & .748$^\star$  & .738$^\star$  & .870  &  & &   \\

\bottomrule
 \end{tabular}}
\caption{Downbeat tracking results using an 8-fold cross validation, where $^\star$ denotes statistical significance compared to TCN (baseline), and $^\dagger$ denotes the predictions made by madmom \cite{bock2016madmom}.}
\label{table:downbeat}
\end{table}

\begin{table}[t]
\centering
\resizebox{\columnwidth}{!}{
\begin{tabular}{l|ccc|ccc} 
\toprule
 & \multicolumn{3}{c|}{{Beat}} & \multicolumn{3}{c}{{Downbeat}} \\
 & F1 & CMLt & AMLt & F1 & CMLt & AMLt  \\
 \midrule
B{\"o}ck et al. \cite{bock2020deconstruct} & .885 & .813 & .931 & .672~~ & .640~~ & .832~~ \\ 
Transformer   & .853 & .741 & .887 & .667~~ & .617~~ & .843~~ \\
TCN (baseline)& .879 & .802 & .911 & .702~~ & .660~~ & .859~~ \\
SpecTNT       & .883 & .809 & .906 & .745$^\star$ & .696$^\star$ & .875$^\star$   \\
SpecTNT-TCN & .887 & .812 & .920 & .756$^\star$ & .715$^\star$ & .881$^\star$  \\

\bottomrule
 \end{tabular}}
\caption{Beat and downbeat tracking result on (test-only) \emph{GTZAN}, where $^\star$ denotes statistical significance compared to TCN (baseline).}
\label{table:gtzan}
\end{table}

From the downbeat tracking results in Table~\ref{table:downbeat}, similar observations can also be made. The performance difference between TCN and SpecTNT is more obvious. Such results are in line with our research motivation that SpecTNT can help capture harmonic and timbral information and improve downbeat tracking. The improvement varies across different datasets. Our proposed models perform better than the baseline by a wider margin for non-pop-centric datasets: \emph{Ballroom} and \emph{Hainsworth}. 
In particular, the performance difference between SpecTNT (or SpecTNT-TCN) and TCN are statistically significant in terms of \emph{F1} and \emph{CMLt}. As mentioned before, SpecTNT may sacrifice its beat tracking performance for downbeat, as the performance gain of downbeat is larger. Qualitatively, we also found SpecTNT can avoid quite a few phase errors as compared to TCN.


Lastly, we present the results of \emph{GTZAN} in Table~\ref{table:gtzan}. In addition to the models presented above, we also include the results from a regular Transformer \cite{won2021transformer}, which contains only the temporal encoder \cite{devlin2018bert}. As it is non-hierarchical and without the spectral encoder, we treat the comparison as the ablation study. It is clear that the regular Transformer does not work well in this case, most likely because the training data is insufficient. Whereas SpecTNT's design can handle this well by leveraging spectral encoders in a stacked architecture. Moreover, we once again see similar results to previous experiments when comparing among SpecTNT, SpecTNT-TCN, and TCN. 

\section{Conclusion}
\label{sec:conclusion}

In this paper, we present a novel Transformer-based model, SpecTNT, to model beats and downbeats from audio signals. 
The design of spectral and temporal encoders enables a promising DNN approach, showing SOTA results in downbeats tracking. 

For future work, we will try to mitigate the sub-optimal issue (see Section \ref{sec:discussion}) with better solutions, as we expect it is currently the limitation for SpecTNT to improve beat tracking. 
We also plan to study the possibility to include additionally relevant MIR tasks (e.g., music structure \cite{wang2022tocatch}, chords, and melody) into a unified multi-task learning framework and model them jointly with SpecTNT.



\bibliographystyle{IEEEbib}
\bibliography{citations}


\end{document}